\documentclass[conference]{IEEEtran}

\IEEEoverridecommandlockouts
\usepackage{array, graphicx, float}
\usepackage{stix}
\usepackage{amsmath}
\usepackage{ragged2e}
\usepackage{titlesec}
\usepackage{booktabs}
\usepackage{fancyhdr}
\usepackage{soul}
\usepackage[english]{babel}
\usepackage[style=ieee, backend=biber, minbibnames=2, maxbibnames=2]{biblatex}

\usepackage{csquotes}
\usepackage{rotating}
\usepackage[normalem]{ulem}
\usepackage{enumitem}
\usepackage{textgreek}
\usepackage{listings}
\usepackage{caption}

\usepackage{tabularx}

\usepackage{xcolor}
\begin{document}
\captionsetup[figure]{name={Fig.}}

\title{Phase Correction using Deep Learning for Satellite-to-Ground CV-QKD\\
}

  

\author{Nathan~K.~Long$^*$,~Robert~Malaney$^*$,~Kenneth~J.~Grant$^\dagger$%
\thanks{$^*$Nathan K. Long and Robert Malaney are with the School of Electrical Engineering and Telecommunications, University of New South Wales, Kensington NSW 2052, Australia.}%
\thanks{$^\dagger$Kenneth J. Grant is with the Sensors and Effectors Division of Defence Science and Technology Group, Edinburgh SA 5111, Australia.}%
}

\IEEEaftertitletext{\vspace{-1\baselineskip}}


\maketitle


\begin{abstract}
Coherent measurement of quantum signals used for continuous-variable (CV) quantum key distribution (QKD) across satellite-to-ground channels requires compensation of phase wavefront distortions caused by atmospheric turbulence. One compensation technique involves multiplexing classical reference pulses (RPs) and the quantum signal, with direct phase measurements on the RPs then used to modulate a real local oscillator (RLO) on the ground - a solution that also removes some known attacks on CV-QKD. However, this is a cumbersome task in practice - requiring substantial complexity in equipment requirements and deployment. As an alternative to this traditional practice, here we introduce a new method for estimating phase corrections for an RLO by using only intensity measurements from RPs as input to a convolutional neural network, mitigating completely the necessity to measure phase wavefronts directly. Conventional wisdom dictates such an approach would likely be fruitless. However, we show that the phase correction accuracy needed to provide for non-zero secure key rates through satellite-to-ground channels is achieved by our intensity-only measurements. Our work shows, for the first time, how artificial intelligence algorithms can replace phase-measuring equipment in the context of CV-QKD delivered from space, thereby delivering an alternate deployment paradigm for this global quantum-communication application.
\end{abstract}


\section{Introduction}

Global quantum key distribution (QKD) could be established by connecting a network of satellite and ground stations via atmospheric free-space optical (FSO) channels~\cite{Bonato2009,Liao2018,Hosseinidehaj2019,Djordjevic2020,Dequal2021}. FSO transmission offers several advantages over optical fibre, including lower attenuation and higher bit-rate communications~\cite{Bloom2003, Malik2015}. However, FSO communications are susceptible to several issues, most important of which being signal loss and signal fading due to distortions caused by atmospheric turbulence~\cite{Zhu2002,Carrasco-Casado2020}. These issues are of critical importance to the implementation of continuous-variable QKD (CV-QKD) between satellites and ground stations - the subject of this work.

Local oscillators (LOs) with known properties have commonly been multiplexed with quantum signals across atmospheric channels to achieve coherent measurements at the receiver. However, transmitted LOs (TLOs) create a security loophole for an eavesdropper (Eve) to intercept and even modify the TLO, rendering the security of the QKD vulnerable (e.g.~\cite{Ma2013,Jouguet2013}). One potential solution to mitigate the vulnerabilities of a TLO is to use a real local oscillator (RLO), also known as a local local oscillator, which is generated at the receiver~\cite{Marie2017,Qi2018,Kish2021}. Coherence between the RLO and quantum signal is achieved by multiplexing reference pulses (RPs) with the quantum signals across a channel (e.g. using a different polarization), such that atmospheric distortion of the RPs can be measured. Wavefront distortion on the RPs is then used to modulate the phase of the RLO to compensate for atmospheric phase drift.

Previous studies into the use of an RLO for CV-QKD include the use of a heralded hybrid linear amplifier to improve signal-to-noise ratios across fibre optic channels~\cite{Li2021}, an investigation into the effects of atmospheric turbulence on secret key rates as a function of satellite zenith angle~\cite{Villasenor2020}, and the development of a detailed noise model for FSO RLO-based CV-QKD~\cite{Kish2021}. Further, several studies into potential attacks on RPs have been undertaken~\cite{Ren2019,Shao2022}.

While it has been proposed that wavefront sensors be used to directly measure phase distortions across atmospheric channels, real-time measurement is restricted by the computational requirements of phase reconstruction algorithms~\cite{Knapek2011}. This current work instead proposes using deep learning to estimate phase wavefront corrections across satellite-to-ground channels with only RP intensity distributions as input. Estimation of wavefront corrections using intensity information on the ground could lead to faster and simpler quantum signal measurements, increasing the practicality of satellite-based CV-QKD.

The open question we investigate here is the following:
Can simple intensity measurements of RPs, transmitted through a satellite-to-ground channel, provide accurate assessments on the phase distortions of RPs - \emph{sufficient to enable non-zero secure key rates for CV-QKD?} Surprisingly, we find the answer in the affirmative. Moreover, we find the secure key rates to be quite significant for a range of channel settings. We believe our results to be the first to show how deep-learning algorithms can replace complex measuring equipment in the context of CV-QKD delivered from space, thereby opening up a new avenue for real-world deployments of this emerging quantum-communication technology.

 The remainder of this paper is as follows:
 A description of turbulent atmospheric satellite-to-ground channel modelling is presented in Section~\ref{sec:channel}. A methodology for estimating phase wavefront corrections across satellite-to-ground channels using only intensity distributions as input to a deep learning model is provided in Section~\ref{sec:cnn}. A secure key rate analyses for CV-QKD is provided in Section~\ref{sec:cvqkd}, and our concluding remarks given in Section~\ref{sec:conc}. 



\section{The Satellite-to-ground Channel} \label{sec:channel}

To model our channel, satellite-to-ground RP laser beam propagation is simulated through varying turbulent atmospheres using a phase screen model based on WavePy~\cite{Beck2016}. This model uses an anisotropic non-Kolmogrov model with a Fresnel split-step propagator. Phase screens represent a turbulent atmospheric volume by assuming that optical beams travel as if in a vacuum until interaction with the screen. The screens then perturb the phase and intensity of the beam when it passes through them, replicating how the beam would be perturbed if it had passed through an equivalent atmospheric volume.

The WavePy model in~\cite{Beck2016} was modified to account for the vertical variation in atmospheric turbulence across various satellite-to-ground channels. The atmosphere was separated into $N$ layers (as described in~\cite{Martin1988}), with varying propagation distances to account for the increase in atmospheric density at lower altitudes. Values for the refractive index structure parameter of turbulence ($C^2_n$) were calculated as in Eq.~(\ref{eq:cn2}) below as a function of altitude ($h$) in m, root mean square wind speed ($\nu_{rms}$) in m/s, and ground-level $C^2_n$ in m$^{-2/3}$, $C^2_n(0)$,
\begin{equation}\label{eq:cn2}
    \begin{split}
        C^2_n(h) &= (0.00594 (\nu_{rms}/27)^2 \ (h \times 10^{-5})^{10} \ \exp(-h/1000) \\ 
        & + 2.7 \times 10^{-16} \ \exp(-h/1500) + C^2_n(0) \exp(-h/100)).
    \end{split}
\end{equation}

\noindent Atmospheric stratification into $N$ slabs in the phase screen model involved an optimization based on the scintillation index ($\sigma^2_I$), where
\begin{equation}\label{eq:sigmai}
    \sigma^2_I = \exp \left[ \frac{0.49 \sigma^2_R}{(1 + 1.11 \sigma^{12/5}_R)^{7/6}} + \frac{0.51 \sigma^2_R}{(1 + 0.69 \sigma^{12/5}_R)^{5/6}} \right] - 1,
\end{equation}

\noindent and as a function of Rytov variance, ($\sigma^2_R$), where
\begin{equation}\label{eq:sigmar}
    \sigma^2_R = 2.25 k^{7/6} \sec^{11/6}(\theta_z) \int_{h_0}^{H} C^2_n (h)(h-h_0)^{5/6} dh.
\end{equation}
Here, $k$ represents the wavenumber, $k = 2\pi/\lambda$. 
\noindent For simplicity in the modelling, the satellite zenith angle ($\theta_z$) was assumed to be $0^\circ$ (as such our results refer to the time when the satellite is directly over the ground station). The entire atmospheric channel parameters used  are outlined in Table~\ref{tab:sim_params}.

\begin{table}[htbp]
    \begin{center}
        \caption{Simulation parameters.}
        \label{tab:sim_params}
        \begin{tabular}{|l|c|}
            \hline
            \textbf{Parameter} & \textbf{Value} \\ \hline 
            Satellite altitude ($H$) & 300 - {500km} \\ \hline
            Ground station altitude ($h_0$) & {0.00} - 2.00km \\ \hline
            Ground-level $C^2_n$ $\left(C^2_n(0)\right)$ & {$1.70 \times 10^{-15}$} - $10^{-14}$ m$^{-2/3}$  \\ \hline
            Outer scale of turbulence ($L_0$) & 5.00m \\ \hline
            Inner scale of turbulence ($l_0$) & 0.025m \\ \hline
            Root mean square wind speed ($\nu_{rms}$) & 21.0m/s \\ \hline
            Satellite zenith angle ($\theta_z$) & $0^\circ$ \\ \hline
            Laser wavelength ($\lambda$) & 1550nm \\ \hline
            Beam waist radius ($w_0$) & 0.150m  \\ \hline
            Receiver radius ($R_r$) & 0.750 - {1.25m} \\ \hline
            RP photon number ($n_{ph}$) & 55,000 \\ 
            \hline
        \end{tabular}
    \end{center}
\end{table}

Several satellite-to-ground channels within the ranges outlined in Table~\ref{tab:sim_params} were simulated to assess the robustness of the phase correction methodology. However, a demonstrative channel (termed \textit{channel one}), with $H = 300$km, $h_0 = 2$km, $C^2_n(0)$ = $1.70 \times 10^{-14}$ m$^{-2/3}$, and $R_r = 0.73$m, will be used, in the first instance, as our key illustrative example (additional channel discussed later).

\begin{figure}[htbp] 
    \centering
    \includegraphics[scale=0.75]{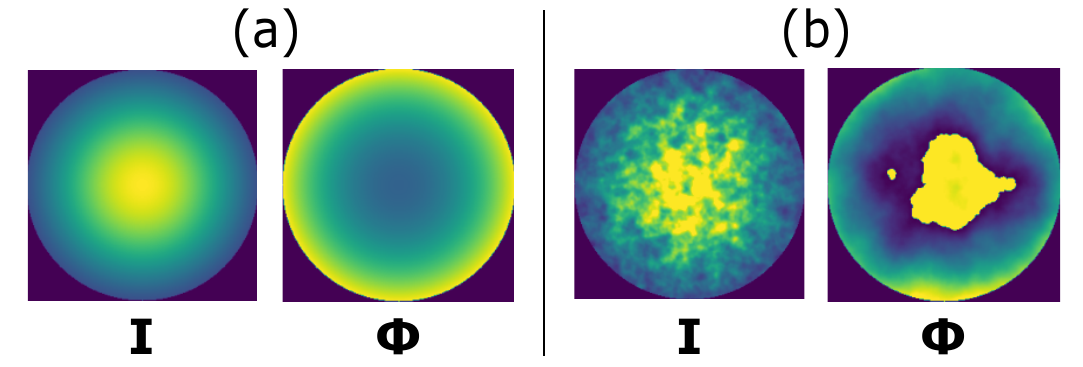}
    \caption{Intensity ($I$) and phase ($\Phi$) distributions after RP transmission through (a) a $298$km vacuum channel and (b) an example instance of {channel one}.}
    \label{fig:rp_dist}
\end{figure}

A beam transmitted across a vacuum (no atmospheric disturbance) diffracts as a function of distance travelled, with a beam travelling through an atmosphere of equivalent distance experiencing equivalent diffraction. Therefore, in order to quantify the phase drift caused by the atmosphere, phase distributions at the ground station were compared to the phase distribution of a beam having travelled across an equivalent distance through a vacuum. The intensity ($I$) and phase distributions ($\Phi$) of a RP after transmission through a vacuum of distance 298km are shown in Fig.~\ref{fig:rp_dist}(a), while example intensity and phase distributions of a random RP after passing through {channel one} are shown in Fig.~\ref{fig:rp_dist}(b).

Some 64,000 instances of RP propagation were simulated for several satellite-to-ground channels within the ranges outlined in Table~\ref{tab:sim_params}. Intensity and phase information was recorded at the receiver for each channel type, then used to train a deep learning model to estimate RLO phase modulation.

\section{Phase Correction Deep Learning Model} \label{sec:cnn}

As stated earlier, we analyse in this work an alternative approach to modulating the RLO phase, whereby only the intensity distributions of the RPs need to be measured at the receiver. Using these measurements, a convolutional neural network (CNN) estimates the phase correction to be applied to the RLO. Encoder-decoder CNNs are designed for image-to-image translation. The encoder maps relationships between input data points to a reduced feature space, then the decoder uses the input feature space to reconstruct a prediction of the output data points. As such, we develop an encoder-decoder CNN model, where the encoder creates a feature space from the received RP intensity distribution, then the decoder constructs an estimated RLO phase wavefront correction for coherent measurement of the quantum signals.

\begin{figure}[htbp] 
    \centering
    \includegraphics[scale=0.6]{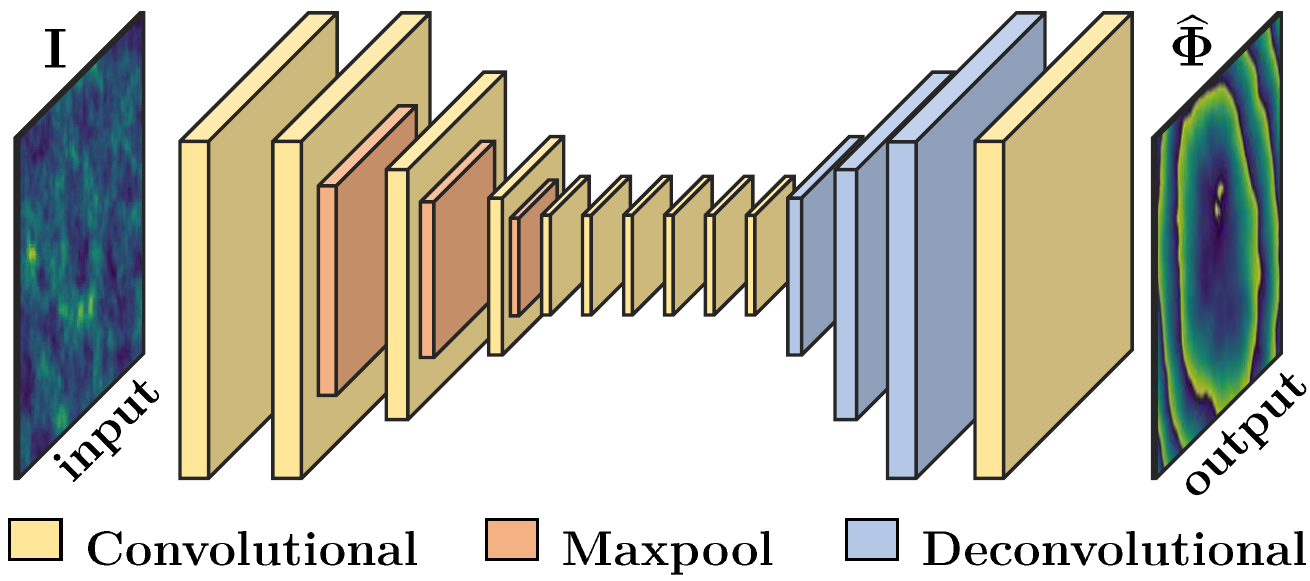}
    \caption{SegNet network layer architecture (further defined in~\cite{Badrinarayanan2017}). Reference pulse intensity distributions ($I$) at the receiver were used as input to the network, which were then mapped to output estimated phase corrections ($\hat{\Phi}$).}
    \label{fig:segnet}
\end{figure}

Fig.~\ref{fig:segnet} outlines the SegNet CNN architecture used to estimate the phase correction (hereby referred to as \textit{the network}), based on the SegNet model described in~\cite{Badrinarayanan2017}, which is comprised of eleven convolutional layers, three maxpool layers, and three deconvolutional layers. The encoder architecture was based on the VGG16 model~\cite{Simonyan2014}, while the decoder inverts the layer sequence of the encoder.

While many image translation networks have been designed to only modify the surface appearance of the same underlying structure (e.g. converting a satellite photo to a map~\cite{Isola2017}), a common underlying structure does not exist between an intensity distribution and phase distribution. Therefore, linear regression to map the final network layer to the predicted phase correction was implemented, without pixel-wise classification or skip connections between the encoder and decoder layers.

The first convolutional layer had a $5\times5$ kernel, while the rest of the layers had a $3\times3$ kernel. Batch normalization was performed on all of the layers, with each using the rectified linear unit activation function. A mean squared error (MSE) loss function was used to optimize the network weights, where the MSE was taken for each pixel then averaged via 
$\frac{\sum^{N_r}_{i=1} \sum^{N_r}_{j=1} (\hat{\phi}_{ij} - \phi_{ij})^2 }{N_r^2}.$
Here, $N_r$ is the number of pixels of the measured RP phase at the receiver, $\phi_{ij}$ is the true phase correction and $\hat{\phi}_{ij}$ is the estimated phase correction at each pixel.

\begin{figure}[htbp]
    \centering
    \includegraphics[scale=0.7]{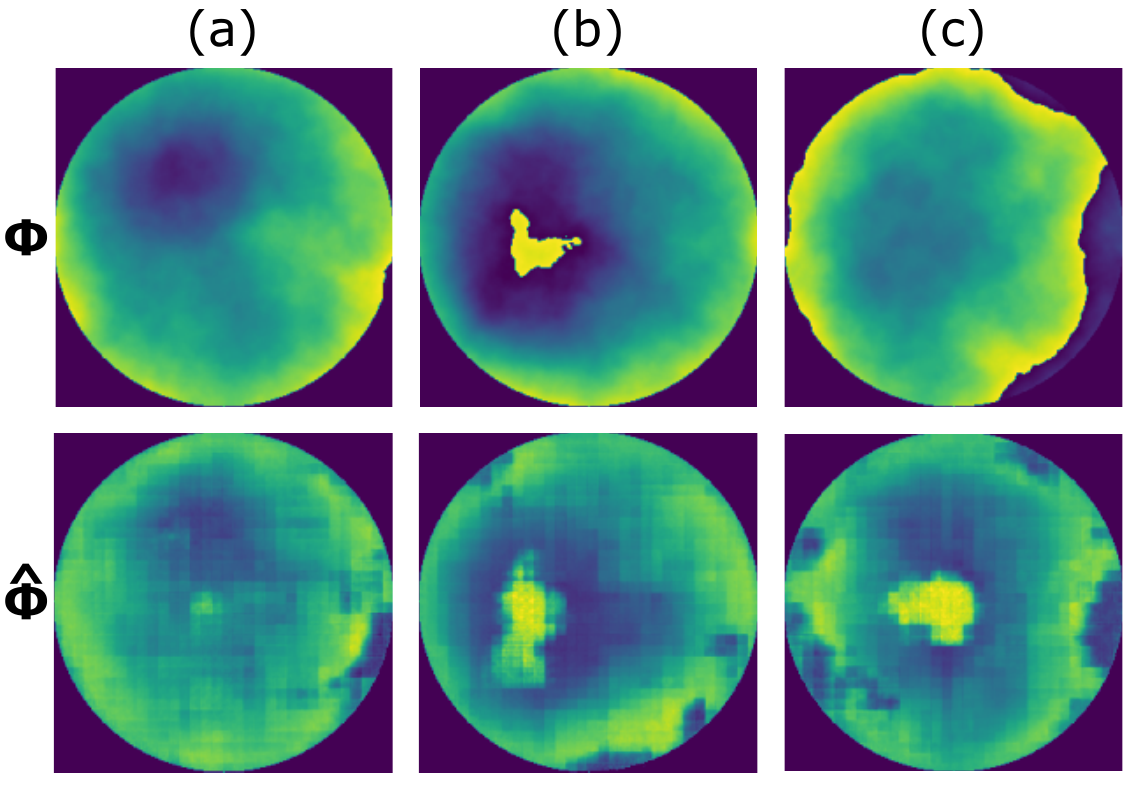}
    \caption{Examples (a)-(c) give RLO phase corrections at the receiver for three independent instances of RP propagation across {channel one}, with true phase corrections ($\Phi$) compared to estimated phase corrections ($\hat{\Phi}$).}
    \label{fig:phi_true_est}
\end{figure}

The 64,000 RP propagation simulations for each channel were split into 57,600 training instances and 6,400 testing instances. Three examples of the CNN phase correction output from the test data {for channel one} are shown in Fig.~\ref{fig:phi_true_est}(a)-(c), where true RLO phase corrections at the receiver (top) are compared with RLO phase correction estimations (bottom).

Considering channel one, the network was generally able to estimate phase gradient trends across the phase correction distributions. For example, higher phase correction values at the edges of the phase distributions were estimated as decreasing towards the center of the distributions, as was present for the true corrections in Fig.~\ref{fig:phi_true_est}. Another example of accurate phase gradient estimation can be seen in Fig.~\ref{fig:phi_true_est}(b), where a phase correction peak was estimated in the center of the distribution, as in the true correction.

The phase gradients in the estimations were found to be more segmented than the true gradients, which could be explained as the optimization function segmenting the output into regional averages to reduce the total MSE between pixels (defining sharp gradients incorrectly would lead to greater MSE values). Further, some output phase corrections estimated false peaks, as in Fig.~\ref{fig:phi_true_est}(c), while others failed to estimate any peaks. However, overall the estimated phase corrections improved coherent efficiencies (defined below) of the RLO. {Note, although not explicitly shown, similar trends were found from the estimation results of the other channel types we use in the  analyses of CV-QKD provided next.}

 We note the prevailing method for recovering RP phase distributions at the receiver is by the use of Zernike polynomials, e.g.~\cite{Villasenor2020,Kish2021}, which required the phase wavefront of the RPs to be measured in order to parameterize the Zernike coefficients. However, wavefront sensor performance deteriorates when the signals encounter strong turbulence, while also requiring computationally intensive algorithms for phase reconstruction~\cite{Knapek2011}, reducing the practicality of real-time implementation. We also note that \cite{Wang2021} used an encoder-decoder CNN architecture to estimate phase distortion across an atmospheric channel with intensity distributions as input. However, the calculations in \cite{Wang2021} focused on single-photon OAM mode recovery for high-dimensional discrete variable QKD across a horizontal atmospheric channel, rather than the phase-correction methodology for satellite-to-ground CV-QKD that is presented here. OAM CV-QKD, although in principle possessing higher throughput advantages, is very difficult to implement in practice, whereas the standard CV-QKD we investigate here is considered mainstream and easier to implement.

\section{Connections to CV-QKD } \label{sec:cvqkd}

 In connecting to CV-QKD, we must apply our phase distortion corrections to the RLO used to measure the incoming quantum signal.
 \subsection{Coherent Efficiency}
 The coherent efficiency ($\gamma$) is used as a singular metric to describe the wavefront distortion. This is derived by comparing the difference between the electric fields of the RLO ($E_{RLO}$) and the RPs ($E_{RP}$) via the following relation,
\begin{equation}\label{eq:coh_eff}
    \gamma = \frac{|\frac{1}{2} \iint_{\mathcal{D}_R}(E^*_{RLO} E_{RP} + E_{RLO}E^*_{RP}) \ ds|^2}{\iint_{\mathcal{D}_R} |E_{RLO}|^2 \ ds \iint_{\mathcal{D}_R} |E_{RP}|^2 \ ds}.
\end{equation}

\noindent The RLO is assumed to be prepared as a beam only affected by diffraction before the phase wavefront correction has been applied. Correcting an RLO phase wavefront increases $\gamma$, where a value of $\gamma = 1$ would indicate a perfect correction of the wavefront. It is assumed that the wavefront distortion is equivalent for the quantum signal and RPs.

\subsection{CV-QKD}
Beyond atmospheric attenuation, absorption, and scintillation, optical signals are affected by channel excess noise ($\xi_{ch}$) and detector noise ($\xi_{det}$), both contributing to key rate reduction for CV-QKD.
$\xi_{det}$ was calculated as a function of $\gamma$, electronic noise ($\xi_{el}$), and detector efficiency ($\eta_{det}$) via,
\begin{equation}\label{eq:xi_det}
    \xi_{det} = \frac{((1-\gamma) + \xi_{el}) \eta_{det}}{\gamma}.
\end{equation}
\noindent The transmissivity ($T$) of a channel further affects key rates for satellite-to-ground CV-QKD~\cite{Hosseinidehaj2019}. This parameter is defined as $T =(P_R/P_T) \eta_{det}$, where $P_T$ and $P_R$ represent the power at the transmitter and receiver, respectively.
Rather than using ensemble-average quantum state information to analyse key rates for satellite-to-ground CV-QKD, a combined \textit{effective transmissivity} and \textit{effective excess noise} term ($T_f \xi_f$) can be used for a non-fluctuating channel (described in~\cite{Villasenor2020}), calculated as,
   $ T_f \xi_f = \xi_{ch} \langle T\rangle + \langle \xi_{det} \rangle + \left(\langle T \rangle - \langle \sqrt{T}\rangle^2\right) V_{mod}.$
Average transmissivity ($\langle T \rangle$) and average detector noise ($\langle \xi_{det} \rangle \approx 1.05$SNU) values were determined for the RPs for each channel type simulated. Using these parameters, the shared information between Alice and Bob ($I_{AB}$) can be calculated as,
\begin{equation}\label{eq:inf_ab}
    I_{AB} = \frac{1}{2} \log_2(1 + \frac{T_f V_{mod}}{1 + T_f \xi_f}).
\end{equation}

\noindent Table~\ref{tab:noise_params} outlines the noise and efficiency values used to analyse key rates in this work.

\begin{table}[htbp]
    \begin{center}
        \caption{Noise parameters.}
        \label{tab:noise_params}
        \begin{tabular}{|l|c|}
            \hline
            \textbf{Parameter} & \textbf{Value} \\ \hline 
            Detector efficiency ($\eta_{det}$) & 95$\%$~\cite{Kish2021}  \\ \hline 
            Reverse reconciliation efficiency ($\beta$) & 95$\%$~\cite{Kish2021} \\ \hline
            Detector electronic noise ($\xi_{el}$) & 0.010 SNU~\cite{Wang2018} \\ \hline
            Channel noise ($\xi_{ch}$) & 0.0172~\cite{Kish2021} \\ 
            \hline
        \end{tabular}
    \end{center}
\end{table}

We adopt the entanglement-based GG02 protocol~\cite{Grosshans2002} in order to calculate key rates before and after the phase correction has been applied. Homodyne detection is used, as this is known to achieve higher key rates than heterodyne detection for low transmissivity values~\cite{Wang2018}. As such, the covariance matrix ($M_{AB}$) describing the quantum state ($\rho$) at an untrusted ground station is given by,
\begin{align}\label{eq:cov_mat}
        M_{AB} &= \begin{pmatrix}a\mathbb{1} & c\sigma_z\\c\sigma_z & b\mathbb{1}\end{pmatrix} \\
        &= \begin{pmatrix}(V_{mod}+1)\mathbb{1} & \sqrt{T_f(V^2_{mod} + 2 V_{mod})} \sigma_z \\\sqrt{T_f(V^2_{mod} + 2 V_{mod})} \sigma_z & (T_f V_{mod} + 1 + T_f \xi_f) \mathbb{1} \end{pmatrix}, \notag
\end{align}

\noindent where $\mathbb{1}$ represents the identity matrix and $\sigma_z$ represents the Pauli-z matrix (of dimension $2\times2$). 

As we shall see, given the atmospheric turbulence, the effects of trust in the detector noise becomes important. A detector is considered trusted if Eve cannot control the imperfection of Bob's detector~\cite{Lin2020}. For the GG02 CV-QKD protocol with reverse reconciliation, trusted detector noise only affects the shared information between Alice and Bob, not the Holevo information between Bob and Eve, whereas effective transmissivity and excess noise affects the Holevo information when detector noise is untrusted~\cite{Laudenbach2018}. Therefore, for a trusted detector, $b = T_f V_{mod} + 1 + \xi_{ch}$ (note the difference from that explicitly given in Eq.~(\ref{eq:cov_mat}) for the untrusted model).

The upper bound on the amount of information that Eve can determine about a quantum state is known as Holevo information~\cite{Laudenbach2018}. Holevo information is calculated as $\chi_{BE} = S_{AB} - S_{A|B}$, where $S_{AB}$ is the von Neumann entropy of the state accessible to Eve and $S_{A|B}$ is the entropy after homodyne measurement by Bob. Here, the subscripts $A$, $B$, and $E$ represent Alice, Bob, and Eve, respectively. Assuming that Eve has a purification of Alice and Bob\textquoteright s shared quantum state ($\rho_{AB}$), the von Neumman entropy ($S$) is $S(\rho_{AB}) = \sum_i g(\nu_i)$, using the function $g(x)$ given by,
\begin{equation}\label{eq:gx}
    g(x) = \frac{x+1}{2} \log_2 \left(\frac{x+1}{2}\right) - \frac{x-1}{2} \log_2 \left(\frac{x-1}{2}\right).
\end{equation}

\noindent Here, $\nu_{i}$ are three symplectic eigenvalues (defined as the modulus of any eigenvalue) of the following matrices. The first two, $\nu_{1,2}$, are the symplectic eigenvalues of $M_{AB}$, and are derived as $\nu_{1,2} = \frac{1}{2}(z \pm [b - a])$, where $z = ([a+b]^2 - 4c^2)^{1/2}$~\cite{Weedbrook2012}. The the third, $\nu_{3}$, is the symplectic eigenvalue of the conditional matrix $M_{A|B}$, where
\begin{align}\label{eq:cov_mat_meas}
        \tilde{M}_{A|B} &= i\Omega M_{A|B} \\
        &= i\begin{pmatrix}0 & 1 \\ -1 & 0\end{pmatrix} \begin{pmatrix}a - \frac{c^2}{b} & 0 \\ 0 & a\end{pmatrix} = i\begin{pmatrix}0 & a \\ -a + \frac{c^2}{b} & 0\end{pmatrix}, \notag
\end{align}

\noindent which results in $\nu_{3} = (a[a - \frac{c^2}{b}])^{1/2}$~\cite{Laudenbach2018}. The Holevo information is then calculated as $\chi_{BE} = g(\nu_1) + g(\nu_2) - g(\nu_3)$.
Finally, the secure key rates, $R_{sec}$, in bits/pulse can be calculated for the simulated satellite-to-ground channels, where $R_{sec} = \beta I_{AB} - \chi_{BE}$, with reverse reconciliation efficiency $\beta$. We note this rate derivation assumes the asymptotic limit for the quantum signal under the assumption of a collective attack.

\subsection{Results}
\begin{figure}[htbp] 
    \centering
    \includegraphics[scale=0.7]{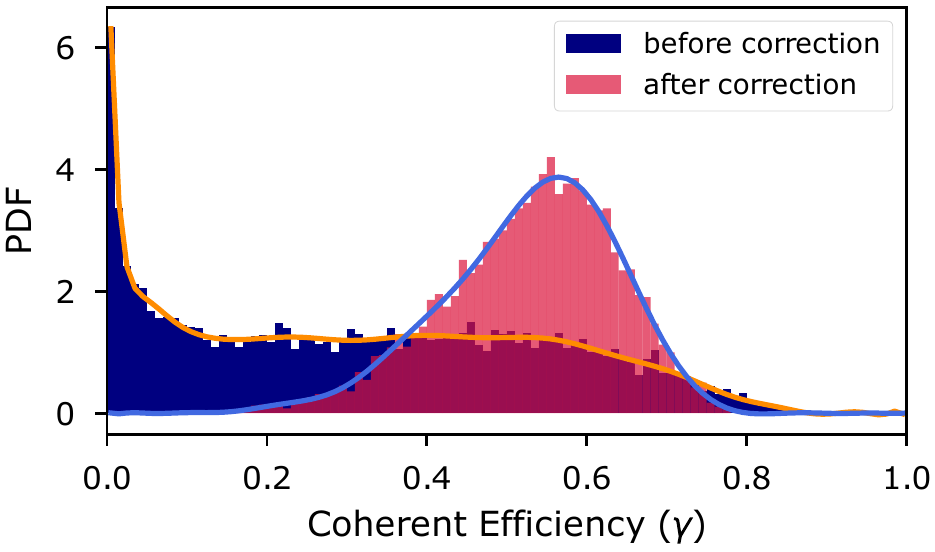}
    \caption{Coherent efficiency probability density distributions, before phase correction was applied (purple) and after phase correction was applied (red), for {channel one}.}
    \label{fig:gamma_pdf}
\end{figure}

The PDFs for coherent efficiency, both before and after phase drift correction, are given in Fig.~\ref{fig:gamma_pdf} for the 6,400 RP simulations used as test data for the CNN across {channel one}.

Our results in Fig.~\ref{fig:gamma_pdf} show that before the phase correction was applied, coherent efficiencies were skewed towards zero, where a value of zero indicates no coherence between RP phase distribution at the receiver and the RLO phase distribution (used for coherent measurement of the quantum signal). However, once the estimated phase correction was applied to the RLO, a mean coherent efficiency of approximately 0.530 was achieved. The phase corrections increased the shared information between Alice and Bob from $I_{AB} = 1.910 \times 10^{-5}$ bits/pulse to $I_{AB} = 0.284$ bits/pulse. As such, after the corrections were applied, quantum signals were able to be measured with greater coherence at the receiver, leading to an increase in shared information between Alice and Bob.

As mentioned earlier, a broad range of channel conditions were tested in order to assess the flexibility of the phase correction methodology across different satellite-to-ground channels (as per Table~\ref{tab:sim_params}). Due to space limitations we cannot show all these results. However, we illustrate some results using three channels representing some of the limits of the channel parameter ranges tested, as defined in Table~\ref{tab:ch_params}. Our conclusions drawn from these channels are consistent with the other results not shown. Compared to channel one, the refractive index structure parameter was weakened across \textit{channel two} by lowering $C^2_n(0)$, while the channel distance was lengthened in \textit{channel three} by increasing the altitude of the satellite. A new network was trained for each channel using 64,000 independent simulation instances, using the same architecture outlined in Section~\ref{sec:cnn}. Channels one and two had an average transmissivity of $\langle T \rangle \approx 0.71$, whereas the receiver radius was increased for channel three to achieve an average transmissivity of $\langle T \rangle \approx 0.83$ to compensate for the lower coherent efficiencies attained from the phase corrections for the longer channel.

\begin{table}[htbp]
    \begin{center}
        \caption{Channel variables.}
        \label{tab:ch_params}
        \begin{tabular}{|l|c|c|c|}
            \hline
            \textbf{Parameter} & \textbf{Channel One} & \textbf{Channel Two} & \textbf{Channel Three} \\ \hline 
            $H$ [km] & 300 & 300 & 500 \\ \hline
            $C^2_n(0)$ [m$^{-2/3}$] & $1.7 \times 10^{-14}$ & $1.7 \times 10^{-15}$ & $1.7 \times 10^{-14}$  \\ \hline
            $R_r$ [m] & 0.75 & 0.75 & 1.50 \\ \hline
        \end{tabular}
    \end{center}
\end{table}

\begin{figure}[htbp] 
    \centering
    \includegraphics[trim={0.7cm 0.5cm 0 0}, scale=0.8]{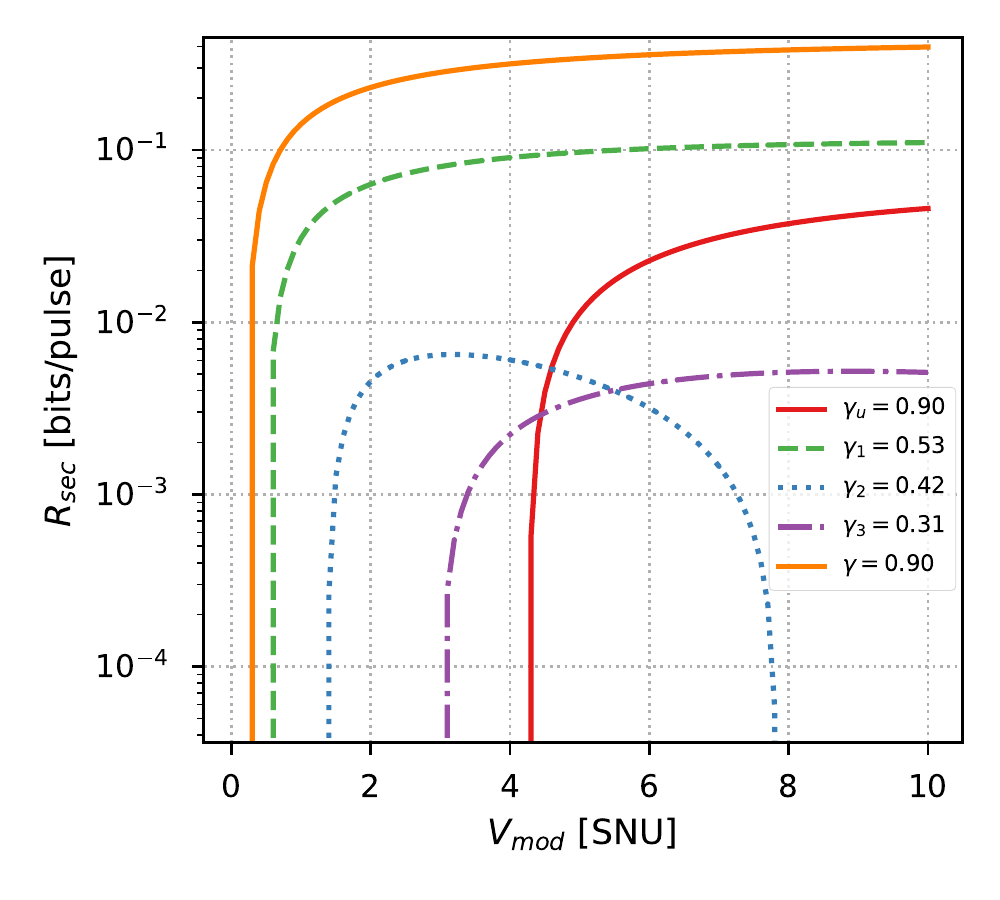}
    \caption{Secret key rate versus $V_{mod}$, for different coherent efficiency values. Results after phase correction are shown for channel one in the dashed green line ($\gamma_1 \approx 0.53$), channel two in the dotted line ($\gamma_2 \approx 0.42$), and channel three in the dash-dotted purple line ($\gamma_3 \approx 0.31$), while results for untrusted and trusted detector noise with $\gamma = 0.90$ are plotted in the red and orange lines, respectively.}
    \label{fig:skr_trust}
\end{figure}

Key rates after RLO wavefront correction across the three channels outlined in Table~\ref{tab:ch_params} are given as a function of $V_{mod}$ and coherent efficiency in Fig.~\ref{fig:skr_trust}. Wavefront corrections for channel one resulted in the highest key rate of the simulated channels, with a coherent efficiency of $\gamma_1 \approx 0.53$ resulting in a maximum key rate of 0.112 bits/pulse. The corrections for channel two produced a maximum key rate of 0.007 bits/pulse at $V_{mod} = 3.2$ SNU, with a coherent efficiency of $\gamma_2 \approx 0.42$, while the phase corrections for channel three led to a maximum key rate of 0.004 bits/pulse, with a coherent efficiency of $\gamma_3 \approx 0.31$. The resulting key rates indicate that channels with weaker turbulence actually result in worse wavefront correction performance when using the methodology outlined in this work, as does increasing the satellite altitude.

Further, results for a hypothetical coherent efficiency of $\gamma \approx 0.90$ are shown for channels with both untrusted ($\gamma_u$) and trusted ($\gamma_t$) detector noise, given $\langle T \rangle \approx 0.71$. Key rates were found to vary greatly depending on whether the detector noise was trusted or untrusted, primarily due to the influence of coherent efficiency on detector noise (see Eq.~\ref{eq:xi_det}). Shown in Fig.~\ref{fig:skr_trust}, an untrusted detector required a coherent efficiency of $\gamma_u \approx 0.90$ to achieve positive key rates, as a result of the trade-off between $I_{AB}$ and $\chi_{BE}$. However, positive key rates were attained for coherent efficiencies above $\gamma \approx 0.42$ for a trusted detector (for $\langle T \rangle \approx 0.71$), when Eve cannot manipulate the detector noise, thus there is less Holevo information between Eve and Bob.

It was found that increasing $V_{mod}$ up to $V_{mod} = 10$ SNU generally led to higher key rates for each of the channels tested, except for channel two which was approaching the limit of achieving non-zero key rates (at $\gamma_2 \approx 0.42$). However, a similar trend was found for each of the other channels tested, where key rates would rapidly increase as $V_{mod}$ increased before reaching an asymptote, reducing the advantage of continually encoding higher $V_{mod}$ values on the quantum states.

A network was then trained to estimate phase corrections for channels of varying $C^2_n(0)$ (across the range in Table~\ref{tab:sim_params}) with $R_r = 0.75$m, achieving an average coherent efficiency of $0.49$ and maximum key rate of 0.072 bits/pulse. As such, given enough training data, it is assumed that a network would be able to output phase corrections for a wide range of channel conditions.

We should be clear: not all satellite-to-ground channel settings we investigate lead to a non-zero key rates. Example settings where $R_{sec}=0$ include channels with a ground station altitude of $h_0 =$ 0km, where the refractive index structure parameter of turbulence is strongest, leading to greater phase wavefront distortions. Of course, any channel can eventually be made to have a non-zero $R_{sec}$ by setting the receiver aperture large enough, but we have not explored that further as large ground-based telescopes obviously increase costs substantially. Finally, global CV-QKD has been our principal interest in this work; as such, we have not explored our setup in the context of terrestrial-only horizontal channels. However, a comparison of effective air-masses between horizontal channels and satellite channels leads us to surmise that for horizontal channels up to 100km, non-zero $R_{sec}$ values will be found for the receiver apertures we have investigated here. That is, we believe our method of intensity-only phase correction will play a substantial role in many terrestrial CV-QKD implementations.

\section{Conclusion} \label{sec:conc}


A convolutional neural network was designed to estimate the correction required to modulate the RLO phase using only the RP intensity distributions at the ground receiver as input for satellite-to-ground channels. We found that the estimated phase wavefront corrections increased coherent efficiencies, resulting in positive key rates for several satellite-to-ground channels when the detector noise was trusted, without the need to measure the wavefront directly. Successful implementation of the proposed method would reduce complexity and increase the feasibility of a global satellite-based CV-QKD network.


\printbibliography



\end{document}